%% file: paper.tex
\def\submission{1}

\documentclass[sigplan,10pt]{acmart}
\renewcommand\footnotetextcopyrightpermission[1]{}
\usepackage{tikz}
\usepackage[]{siunitx}
\usepackage{booktabs} 
\usepackage[linesnumbered,ruled,vlined]{algorithm2e}
\usepackage{xspace}
\usepackage{pifont}
\usepackage{array}
\usepackage{listings}
\usepackage{url}
\usepackage{multirow}
\usepackage{enumitem}
\usepackage{xcolor}
\usepackage{makecell}
\usepackage{comment}
\usepackage{natbib}

\microtypecontext{spacing=nonfrench}

\SetKwComment{Comment}{$\triangleright$\ }{}

\SetAlFnt{\footnotesize}

\newcommand{\sys}{\texttt{Blizzard}\xspace}
\newcommand{\sysrep}{\texttt{liblogrep}\xspace}
\newcommand{\syslib}{\texttt{libds}\xspace}

\if\submission 1
\newcommand\todo[1]{\textcolor{red}{#1}}
\newcommand\ada[1]{\textcolor{blue}{#1}}
\newcommand\pf[1]{\textcolor{green}{#1}}
\else
\newcommand\todo[1]{}
\newcommand\ada[1]{}
\newcommand\pf[1]{}
\fi

\newenvironment{tightitemize}%
 {\begin{list}{$\bullet$}{%
 		\setlength{\leftmargin}{10pt}
        \setlength{\itemsep}{1pt}%
        \setlength{\parsep}{1pt}%
        \setlength{\topsep}{1pt}%
        \setlength{\parskip}{1pt}%
        }%
 }%
{\end{list}}

\definecolor{codegreen}{rgb}{0,0.6,0}
\definecolor{codegray}{rgb}{0.5,0.5,0.5}

\lstdefinestyle{mystyle}{
    commentstyle=\color{codegreen},
    numberstyle=\tiny\color{codegray},
    basicstyle=\ttfamily\footnotesize,
    breakatwhitespace=false,    
    breaklines=true,    
    captionpos=b,    
    keepspaces=true,    
    numbers=left,    
    numbersep=5pt,    
    showspaces=false,
    showstringspaces=false,
    showtabs=false,
    tabsize=2
}

\definecolor{codegreen}{rgb}{0,0.6,0}
\definecolor{codegray}{rgb}{0.5,0.5,0.5}

\begin{document}
\date{}
\title{Blizzard: Adding True Persistence to Main Memory Data Structures}

\author{Pradeep Fernando}
\affiliation{%
  \institution{Georgia Institute of Technology}
  \city{Atlanta}
  \country{USA}
}

\author{Daniel Zahka}
\affiliation{%
  \institution{Georgia Institute of Technology}
  \city{Atlanta}
  \country{USA}
}

\author{Subramanya R. Dulloor}
\affiliation{%
  \institution{Kumo.AI}
  \city{Mountain View}
  \country{USA}
}

\author{Amitabha Roy}
\affiliation{%
  \institution{Kumo.AI}
  \city{Mountain View}
  \country{USA}
}

\author{Ada Gavrilovska}
\affiliation{%
  \institution{Georgia Institute of Technology}
  \city{Atlanta}
  \country{USA}
}


\input{abstract}


\settopmatter{printfolios=true,printacmref=false}
\maketitle

\input{introduction.tex}
\input{motivation.tex}

\input{overview.tex}

\input{interface.tex}

\input{replication.tex}
\input{execution.tex}

\input{impl.tex}
\input{evaluation.tex}

\input{relatedwork.tex}

\bibliographystyle{ACM-Reference-Format}
\bibliography{refs}

\end{document}

%% file: abstract.tex
\begin{abstract}

Persistent memory (PMEM) devices present an opportunity to retain the
flexibility of main memory data structures and algorithms, but augment
them with reliability and persistence. The challenge in doing this is
to combine replication (for reliability) and failure atomicity (for
persistence) with concurrency (for fully utilizing persistent memory
bandwidth). These requirements are at odds due to the sequential
nature of replicating a log of updates versus concurrent updates that
are necessary for fully leveraging the path from CPU to memory. We
present \sys~– a fault-tolerant, PMEM-optimized persistent
programming runtime. \sys addresses the fundamental tradeoff by
combining (1) a coupled operations log that permits tight integration
of a PMEM-specialized user-level replication stack with a PMEM-based
persistence stack, and (2) explicit control over the commutativity
among concurrent operations. We demonstrate the generality and
potential of \sys with three illustrative applications with very
different data structure requirements for their persistent state.
These use cases demonstrate that with \sys, 
PMEM native data structures can deliver up to $3.6\times$
performance benefit over the alternative purpose-build
persistent application runtimes, while being simpler and safer (by
providing failure atomicity {\em and} replication). 

\end{abstract}

%% file: introduction.tex
\section{Introduction}

Persistent memory (PMEM) hardware provides a way around the
cumbersome block storage abstraction. 
We can now access persistent memory at byte granularity directly from the CPU, using the 
same instructions used to access volatile main memory. This in turn means that main memory
data structures using pointers, such as hash tables, graphs and priority
queues, need no longer be volatile.

However, existing enterprise and web applications using such data structures demand
{\bf true persistence}, ensuring the availability of their state even
when persistence media fails. 
The advance in memory hardware still leaves the worrying question of
fault tolerance 
and the right software interface to persistent memory that
hides the complexity of failure atomicity adequately. Combining the
much faster new persistent devices with existing replication
protocols~\cite{raftpaper} shifts much of the end-to-end bottlenecks into the data
transport and copying overheads of the replication stack. It also raises
the question how to couple the execution of the persistence and
replication engines, while providing for concurrent operations, needed
for performance, and maintaining ordering, needed for
correctness. This is non-trivial because the sequential nature of
replicating a log of updates is at odds with the concurrent updates
needed to maximize the performance that can be achieved on the new
persistent media.

In response to these questions, we present {\bf \sys}~-- a fault-tolerant, PMEM-optimized persistent programming
runtime.  \sys is a software stack that lets 
programmers build sophisticated {\em truly persistent data structures as a service}
with only modest software modification requirements. Truly persistent
data structures are 
exposed to client application through an RPC interface, while \sys~
{\em ensures the performance and correctness of the data structures'
access, durability and fault-tolerance operations}.

To achieve this,
\sys relies (1) on a {\em coupled operations log} that permits the tight integration
of a PMEM-specialized user-level replication stack with a PMEM-based
persistence stack, 
and (2) on  {\em 
explicit control over
the commutativity} 
among concurrent operations. 
To realize (1),  
\sys leverages the byte-addressability of new persistent
memories and the direct memory access capabilities of commodity I/O
devices, and provides for end-to-end use of zero-copy and batching.
This allows for sufficient concurrency to keep the network interface
well utilized.
It also allows the persistence and replication
engines to operate on fine granularity, eliding the need for block
storage abstractions, and making it possible to natively support
diverse data structures and their access APIs. 
To ensure ordering and correctness guarantees, \sys relies on  
(2)
by exposing APIs to the programmer (or to upper level software stacks)  to 
specify when updates in the replication log are commutative. \sys utilizes this 
information to relax the strict ordering specified in the replication
log and to execute update operations
in parallel, while 
maintaining serializability guarantees.

We also 
provide a 
small (but growing) set of \sys implementations of popular data
structures,  
that 
allow performant implementations of
applications to benefit from the simplicity
and functionality of main memory data structures, while 
providing performance and true
persistence. 
We demonstrate the
generality and potential of \sys with three illustrative 
applications with very different data structure requirements for their
persistent state:
\begin{tightitemize}
\item a persistent key value store (such as
the PMEM-optimized NoveLSM~\cite{novelsm}) that can be intuitively created with under 100
lines of extensions to an existing 
in-memory unordered map; 
\item a persistent graph database specialized for
streaming updates (modeled after
GraphOne~\cite{graphone}) built with an in-memory, persistent adjacency list, also in
under 100 lines of code; 
and
\item a
modern web-application backend (Lobsters~\cite{lobsters}) built in 600 lines of code with in-memory persistent data
structures that match the application processing requirements
(priority queues and hashmaps) vs.~a specialized relational DB
backend such as Noria~\cite{noria}.
\end{tightitemize}
These use cases 
demonstrate that with \sys, 
PMEM native data structures can deliver up to $3.6\times$
performance benefit over the alternative purpose-build
persistent application runtimes, while being simpler and safer (by
providing failure atomicity {\em and} replication).

In summary, this paper makes the following contributions:

\begin{tightitemize}
  \item We design a PMEM-specialized replication stack
    that addresses the challenges of integrating replication,
    persistence and concurrency when combining PMEM-based persistence
    with high-speed networking. 
  \item We provide the implementation of the \sys system for commodity
    Ethernet networks with DPDK high-speed packet delivery, and 
    Intel Optane DC PMEM, the
    first-generation of directly-attached persistent memories\footnote{We will
      opensource \sys prior to publishing this paper.}. 
  \item 
    The evaluation of \sys with different applications
      demonstrates its generality and flexibility for
      creating, with modest effort, different persistent and fault-tolerant 
    data structures, that can be further combined as needed by
    sophisticated applications, while delivering to applications
    performance benefits and stronger
      availability guarantees. 
\end{tightitemize}


%% file: motivation.tex
\section{Motivation}

PMEM 
allows persistence at byte-granularity and provides 
low latency, high throughput data reads/writes. The systems community has built on new PMEM 
device capabilities and has developed software primitives 
for representing durable application state using intuitive persistent 
memory data-structures. One examples is the Intel-provided Persistent
Memory Development Kid (PMDK)~\cite{pmdk}. 
	PMDK is a popular PMEM programming library that supports PMEM programming primitives such as persistent
        memory allocators and durable transactions. Other recent
        programming systems, such as Pronto~\cite{pronto} and
        Persimmon~\cite{persimmon:osdi20}, provide support for creating
        persistent versions from 
        volatile data structures, including via compiler support and
        dynamic instrumentation to capture the log of
          updates and automate the insertion
         of flush and fence operations needed for crash-consistent
         persistent state. 
However, {\em supporting reliability {\em and} persistence together for in-memory data-structures 
remains an open challenge.}

\begin{figure}[tb]
    \centering
    \includegraphics[width=\linewidth]{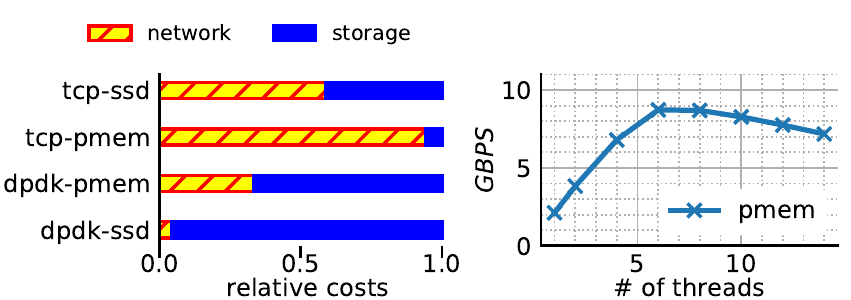}
    \vspace{-4.0ex}
	\caption{\small Relative costs of the network and storage stacks in a replication
	setup with different technologies (left). Impact of concurrent
        accesses on PMEM-achieved throughput for Intel Optane DC
        (right).} 
      \label{plot:latency}
\end{figure}

Reliability and fault-tolerance of persistent data is often achieved
using data replication. 
As shown in the left graph in \autoref{plot:latency}, with existing persistent
technologies such as SSDs, the dominant component of the end-to-end
costs associated with replicated persistent state remains in the
storage layer. For SSD systems, the network-related replication
overheads are easily made negligible with direct use of current commodity
high-speed interconnects, such as with support for DPDK (the Data Plane
Development Kit which provides libraries for fast packet
processing)~\cite{dpdk} or RDMA~\cite{rdma,anuj:socc20}. This has
allowed {\em much of
the past
research to focus their efforts on improving the performance of
existing application storage backends, without considering
opportunities to co-optimize it with network replication.} 

Simply achieving performance by porting or re-designing storage
abstractions for PMEM~\cite{novelsm,novafs} and leveraging existing
replication techniques, but 
without 
supporting persistent
data-structures as a first class citizen.
As also showb in recent research on programming persistent
or disaggregated memory
systems~\cite{pronto,asymvm,fardatastructures,persimmon:osdi20},
{\em real application use diverse and often multiple 
data-structures, requiring performance together with native support
for properties such as persistence and
reliability for all such application state. }

For the new
persistent memory technologies, use of high-speed communications
stacks in the replication solution is critical for 
end-to-end performance, since the
network overheads of traditional kernel-based TCP stacks
dominate. However, simply 
integrating fast networking, specifically userspace networking stacks
such as DPDK or RDMA, with standard PMEM 
data-structures programming libraries~\cite{persimmon:osdi20,pronto}, will fall short on
performance~\cite{mojim}.
{\em Both the network replication {\em and}
the persistence media storage components remain significant contributors
to the end-to-end application performance and seeking reductions in
both of these components remains equally important. }

This is because data and metadata has to move across the network and durable replication 
layers, 
exposing significant overheads due to both the network and the costly memory management 
overheads in the form of allocate, copy and 
garbage-collection, as shown in \S\ref{sec:eval}. Use of well-known
mechanisms such as zero-copy and batching can reduce and amortize
these overheads. Moreover, these techniques promote the concurrent use
of the network transport, opening possibilities to maximize the
concurrent access to the network attached persistent devices. As shown
in the right graph in \autoref{plot:latency}, this is particularly
important for PMEM, whose performance is generally maximized with increased
concurrency.
However, {\em the concurrency required for 
high-throughput in-memory data structure design and low-overhead
persistence, is at odds with the serial execution needed to ensure 
ordering and correctness guarantees for how data structure updates
propagate through the replication stack.}

The solution developed in this paper addresses these gaps by 
building on mechanisms
that integrate best-practice principles for performant
persistence {\em and} replication, namely zero-copy, batching and
concurrency, for platforms combining high-speed networking and
byte-addressable persistent memory technologies, to deliver true
  persistence to native in-memory data structures. We demonstrate this
  for a range of application with different data structure
  requirements.

%% file: overview.tex
\section{Overview}

We propose {\bf \sys}~-- a fault-tolerant, persistent memory programming runtime with
native support for in-memory durable data-structures. \sys
achieves low-latency, high-throughput replication of in-memory state
through use of a PMEM-specialized user-level replication stack.
Key to achieving performance is use of a coupled operations log that
allows for tight integration of the persistence and replications
engines with end-to-end zero-copy, realized by leveraging the
byte-addressability of the PMEM hardware and the direct I/O
capabilities of the network fabric (\S\ref{sec:replication}). 
Performance in the replication path is further enabled by maximizing
concurrency, while continuing to maintain application-specific
ordering and correctness guarantees (\S\ref{sec:execution}). 
The outcome is that \sys provides for reliable and persistent in-memory
data-structures, capable of replacing existing enterprise application
backends while providing improved 
performance, reliability and additional functionality (\S\ref{sec:interface}).

\begin{figure}[tb]   
    \centering
    \includegraphics[width=\linewidth]{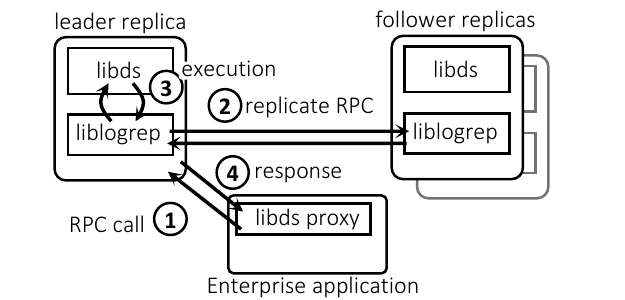}
    \caption{\small \sys consists of three main components. \syslib, \sysrep and Execution protocols
	that couple former two.} 
      \label{fig:overview}
\end{figure}

\sys has two key components, \syslib and \sysrep, shown in \autoref{fig:overview}. \syslib implements a growing number of
persistent memory data structures
modeled after familiar C++ STL library counterparts, and \sysrep handles their persistent state replication
among node replicas. 


{\bf \syslib} is the application developer facing component of \sys. 
	\syslib supports a rich set of commonly used reliable, persistent memory 
	data-structures (e.g. maps, queues, etc).   
	We implement \syslib data-structures using PMDK~\cite{pmdk} -- a popular persistent memory programming library. 
	   The operations to
        create and manipulate 
        each of the data-structures 
        are
	exposed as a network call and
        made available to client-side applications via 
        associated data-structure \texttt{libds proxy}.  
        \syslib allows
        these data-structures to be combined to form complex backend 
	data models required by real enterprise applications. 
	At the heart of \syslib functionality, is the core \sys
        programming APIs supporting {\em RPC-style persistent 
	memory programming}, similar to~\cite{persimmon:osdi20,asymvm}, with
        additional {\em flexibility to describe operations commutativity
        and (re-)ordering requirements}. 


        {\bf \sysrep} is  a PMEM-aware, fast
    log-replication runtime that extends the RAFT log replication protocol. \sysrep 
	log replication is durable and replicates \syslib's operations across replica nodes.
        It carefully
        {\em integrates userspace networking and
        the byte-addressability of PMEMs to 
	realize end-to-end data zero-copy and batching, and to achieve low latency/high-throughput 
	operation replication}.

        \syslib and \sysrep are combined via a common operations log in the \sys execution layer in a
        manner that
        retains the zero-copy benefits while  
        ensuring
	correct end-to-end execution of incoming data-structure
        operations, both within a 
	single node (crash-consistency semantics) 
        and 
	across node replicas (distributed data
        consistency). 
        To increase system throughput, \sys maximizes parallelism
        while preserving ordering constraints 
        and correctness. It does this by {\em integrating in the execution
        layer 
        operations scheduling 
        that allows
	concurrent execution of operations on replica nodes without
        compromising distributed  
	consistency and violating ordering constraints} specified by the \syslib APIs. 

%% file: interface.tex
\section{Creating Data Structures with \sys}
\label{sec:interface}
The core operational model in Blizzard is a client-server one:
application developers write
services that receive remote procedure calls from clients, lookup and manipulate 
persistent state, and then return a response.

The listing below shows the Blizzard API available to programmers; we elide some setup details for reasons
of space and focus on the core APIs.
In order to provide maximum flexibility, when invoking operations on
\sys data structures from a client, the developer specifies the RPC
call as a binary blob. 
\sys takes care of service discovery (primarily, locating the RAFT leader) and sends the server the RPC call.
It returns the response as another binary blob and a \texttt{Status} object detailing whether the RPC call could 
be made successfully. Application level errors, if any, are encoded in the return blob by the application. We distinguish 
between read and update RPCs as separate calls. This is because reads are not replicated but updates need to be.

\begin{lstlisting}[language=C, caption= \sys API.]
	// Client side API
	Status MakeUpdateRPC(const string& request, string* response);
	Status MakeReadRPC(const string& request, string* response);
	
	// Server side API
	class Lock() {
	 public: 
	  virtual void ReleaseLock() = 0;  // Override to release your lock
	}
	void HandleRPC(const string& request, vector<Lock>* delayed_locks, string* response);
	bool Commutes(const string& requestA, const string& requestB);
      \end{lstlisting}

      The server side API 
      {\tt HandleRPC} is implemented by the programmer and executed as a callback 
by \sys. It contains the received request blob and 
returns a response blob after execution. 
For update RPC calls, \sys guarantees that the {\tt HandleRPC} call is executed \emph{after} the call has been
successfully replicated on the RAFT log (\S\ref{sec:replication}). The {\tt HandleRPC} call is executed in the
context of a persistent memory transaction (\S\ref{sec:execution}). Any updates to persistent memory are only
committed \emph{after} {\tt HandleRPC} returns to the \sys runtime.

We expect concurrent invocations of  {\tt HandleRPC} to access a
shared, persistent memory data structure, and to do so 
in a thread-safe manner. 
\syslib provides some thread-safe data structures along with their client-side 
proxies; the programmer is also free to 
roll their own using the PMDK library. In all cases data race safety
 enforcement is the responsibility of
the programmer using their
favorite lock implementation, which 
can be placed in volatile memory. The only requirement is that they
wrap these locks in an object derived from {\tt Lock} in the \sys API
and to \emph{not} release them when executing
the RPC callback. Instead, these must be returned in the vector object provided in the API. \sys releases
all locks \emph{after} updates to persistent memory have been
committed (see discussion in
\S\ref{sec:execution}).

Finally, we expect the programmer to implement a callback to 
determine the commutativity of various RPC calls. 
A programmer declares two RPC calls as commutative if either order of
execution leads to the same result for both calls, 
and therefore,  it is safe if \sys executes them in
different order on different replicas. A simple example of this is
incrementing 
a counter. Commutative RPC calls are executed concurrently,
with better performance then when 
following the sequential execution specified by the replication log.

To make the server side API more concrete, the following listing provides an abbreviated benchmark
from Lobsters~\cite{noria} that we use in this paper. It maintains a hash table, mapping news story identifiers 
to vote counts. The data structure is provided by \sys and for this example we assume it is \emph{not}
thread-safe, requiring the programmer to implement their own locking. The RPC call handler adds a vote to
a story.

\begin{lstlisting}[language=C, caption= \sys sample code for top-K voted entries.]
	blizzard::map<string, int> votes;
	pthread_mutex_lock big_lock; // in volatile memory!
	
	class MyLockWrapper : blizzard::Lock {
	 public:
	   MyLockWrapper(pthread_mutex_lock* lock) 
	    : saved_lock_(lock) {}
	   virtual void ReleaseLock() {
	     pthread_mutex_unlock(saved_lock_);
	   }
	 private:
	  pthread_mutex_lock* saved_lock_; 
	}	
	void HandleRPC(const string& request, vector<Lock>* delayed_locks, string* response) {
	  // Lock everything and stash away the lock for blizzard
	  pthread_mutex_lock(&big_lock);
	  delayed_locks->push_back(MyLockWrapper(&big_lock));
	  votes[StoryId(request)]++;  // Blizzard auto-undo logs the hash bucket!
	}	
	bool Commutes(const string& requestA, const string& requestB) {
	  // Allow vote increments for a story to commute.
	  return true;
	}	
\end{lstlisting}

The most interesting aspect of this example is the decision (by the programmer) to declare all vote increments to commute.
Viewed from the perspective of a single story, increments and reads of the vote count are serialized. From the perspective
of multiple stories however increments are not serialized as the updates to different stories are executed in different orders
on different replicas. This reflects the fact that {\em Blizzard provides strong consistency in terms of state machine replication
in the underlying layers, but allows programmers to relax that ordering for better concurrency}. As we show in the following sections,
we address the performance of RAFT replication while maintaining the serial order of log updates, thereby providing a strongly 
consistent and performant substrate for programmers to build their persistent memory applications as they see fit.

%% file: replication.tex
\section{Replication}
\label{sec:replication}
Replication is necessary in \sys 
to ensure in-memory data structures are
truly available even when the underlying persistent memory fails or
the machine goes down. However, log replication is a synchronous
operation. The latency components
of accessing durable 
storage and network hops to replicas add to the latency of operation
completion.
Although other research has focused on 
network overheads for replication 
protocols~\cite{consensus_box, dare, fast_dc_rpc, curp:nsdi19}, \emph{they have examined replication without persisting any state}.
This is not an accident -- Flash storage comes at a significant latency and throughput cost compared to network
performance. In contrast, \sys is designed ground up for persistent memory that is at least an order of magnitude
faster than Flash.
Even recent works which do use log replication in conjunction with
persistent data structures, do so in a different context: In
Persimmon~\cite{persimmon:osdi20} an operations log of updates is replicated from
a primary in-DRAM node to a local secondary in-PMEM shadow copy. In
AsymNVM~\cite{asymvm} node-local logs of updates to DRAM data
structures are flushed via RDMA operations to a remote PMEM
location. The actual replication to mirror nodes is outside of the
critical path of the persistent memory update, as it can involve
arbitrary persistent devices, including SSDs, using
Zookeepeer. Importantly, this decision exposes applications to
  potential data loss if the original persistent device fails. 
These observations motivate us to focus on the network component of a fully functional
replication stack that \emph{includes} persistence. 

A key building block for \sys is userspace network access.
\sys uses RAFT~\cite{raft} to replicate a durable log of updates to persistent memory data structures. 
We use the Data Plane Development Kit (DPDK~\cite{dpdk}) for fast access to the network from
userspace. For our specific setup, this leads to a 3$\times$ reduction in latency for a single hop on
the network from 28~us down to 8~us.
We then exploit the direct addressability of persistent 
memory to build a high performance PMEM-based replication stack using two simple
principles: {\em zero copy} and {\em batching}.

Copying log entries in various parts of the replication state machine -- from receiving client requests to sending
out copies to replicas -- is expensive. This is even more so since persistent memory is still 
slower than volatile RAM, and its performance, particularly for write
operations, further degrades with increase in thread counts. 
The fact that persistent memory allows one data structure to simply
point to another (rather than indirecting through a block address on Flash)
  makes it possible to avoid copies by reusing the operations log
  entries {\em in place} across all portions of the \sys
  stack. 
  
\begin{figure}
  \includegraphics[keepaspectratio,width=1\linewidth]{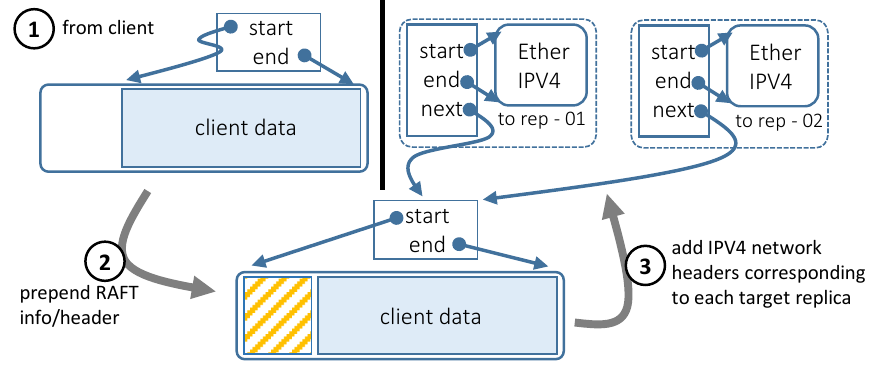}
  \vspace{-4.0ex}
\caption{RAFT log entry in \sys.}
\label{fig:raft_log_entry}
\vspace{-2.0ex}
\end{figure}

Figure~\ref{fig:raft_log_entry} shows how in the RAFT-based
implementation of \sys, log entries are organized 
relative to DPDK's memory
buffers. A DPDK memory buffer holding an incoming request (an Ethernet frame) is placed in an aligned block
of memory together with external metadata pointing to the start and
finish of the block, 
all in persistent memory. We keep the client request in its DPDK memory buffer for its lifetime -- spanning replication
and execution.

To start with, the leader prepends a RAFT control block (with information such as term and index) to the
buffer adjusting the external metadata to compensate.
The leader is now ready to replicate the request. We exploit the fact that
DPDK allows multiple memory buffers to be chained together. To do so it simply creates an Ethernet header for  
\emph{each replica} and  chains the same log entry packet to each of them. It hands off all the headers to DPDK.
The NIC then does the heavy lifting of assembling the Ethernet frames and sending them out. We underline that this 
is only possible because the logs are not in block storage and persistent memory is accessible from all connected
agents, including I/O devices, in the system. In contrast, directly accessing block storage from the NIC involves serious 
complications~\cite{reflex} and this design illustrates how persistent memory can simplify the design of distributed system 
primitives that need persistence.

Although this design lifts most of the load off the CPU,
the 
consensus protocol still represents
an overhead for each log entry. More pertinently, this overhead is sequential, since each log entry needs to be
processed before moving on to the next one. A simple way to further improve performance under load is therefore
to batch process log entries. DPDK already provides an efficient vector interface to receive multiple packets waiting
in the NIC queue. We chain these packet buffers together in userspace and treat them as a single RAFT log entry.
Note that this form of chaining forms a batch from concurrent operations issued
from multiple threads, and need not impact individual operation
latencies.
This
effectively amortizes the CPU cost of running the RAFT protocol state machine over multiple log entries.

\sys's RPC layer, together with the totally ordered semantics of RAFT replication, means that we provide serializability
in terms of the distributed consistency model, if we execute RPC operations in the RAFT log order and read operations as soon as
they are received at the leader. Although, we do not replicate reads, the RPC layer always directs reads to the
leader replica and thus, we provide read your own writes consistency in addition to serializability. We \emph{do not} provide linearizability
as that would require us to replicate reads to ensure that a leader does not become partitioned without realizing it and responds to reads
without taking into account concurrent writes in the majority quorum.
We believe serializability with read your own writes consistency
is an adequately strong distributed consensus model for programmers to
be largely oblivious to replication under the hood.

%% file: execution.tex
\section{Execution Layer}
\label{sec:execution}
The goal of the execution layer is to concurrently execute \emph{committed} operations in the RAFT's execution log.
\sys depends on the application programmer to specify commutativity among operations (Section~\ref{sec:interface}).
The execution layer couples to the replication layer via a set of queues to receive operations and uses flags in
the RAFT log entries to track and update the state of each operation -- replicating, replicated (or replication failed), executing and complete. The most complex part of 
the execution layer is the scheduler that aims to schedule ready operations as soon as possible, while respecting
commutativity.
The actual execution leverages PMDK's persistent memory transaction library to enforce failure atomicity.
We discuss each of these components below. Finally, we also discuss the implications of declaring operations
as commutative and how the programmer can control departure from serial execution order for better performance.

\subsection{Coupling}

\begin{figure*}[ht]
	\centering
	\includegraphics[keepaspectratio,width=1\linewidth]{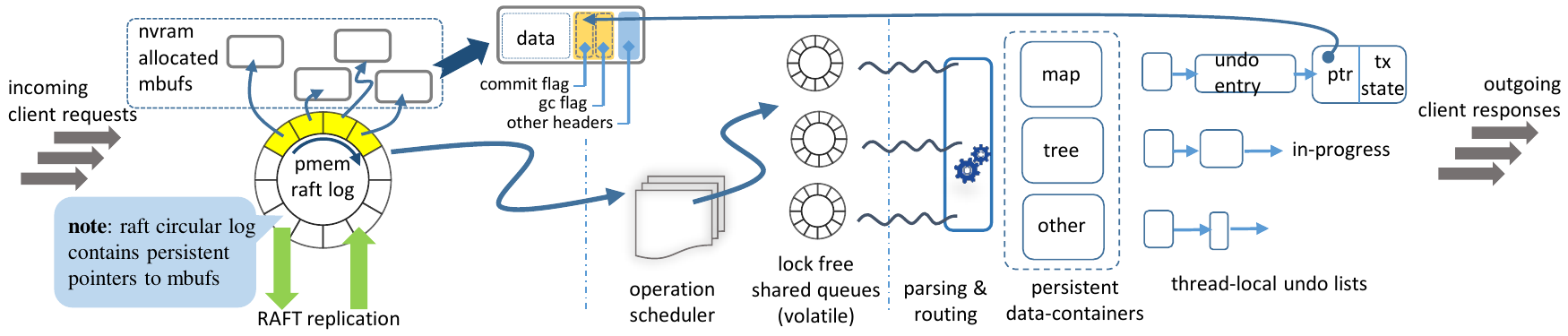}
	\caption{Coupling replication to
          execution in \sys. }
	\label{f:blizzard}
\end{figure*}

Figure~\ref{f:blizzard} illustrates the 
design and interfaces between the replication and execution layers in \sys. Every
 read and write
operation received by the replication layer, is added to a queue ({\tt Q}) of operations, implemented as a persistent circular log. 
Each entry in the circular log is 
a pointer to the actual 
DPDK memory buffer holding the RAFT log entry. Each 
RAFT log entry 
includes 
a set of flags read from and written to by both the execution and
replication layers, 
so as to communicate the state of
the operation. 
These flags are persistent and survive restarts, forming the
basis for recovery. 

All currently executing or ready to execute set ({\tt E}) of operations (a subset of {\tt Q}) are maintained in volatile memory. 
A scheduler picks operations from {\tt Q} and adds them to {\tt E}
when ready to execute. 
Executor threads pick operations from {\tt E} to
execute and update the flags in the RAFT log when execution is complete.  As part of this post completion operation, 
the execution thread marks the operation for garbage collection by setting the \texttt{gc\_flag}. The replication sub-system
uses this information to decide when to garbage collect the RAFT log and move the tail of the persistent circular log
forward.

\subsection{Scheduling}
The scheduler runs as part of the continuous event loop in Blizzard,
executing the scheduling
algorithm
shown in Algorithm~\ref{schedulealgo}.
It considers for execution operations at the head of  {\tt
  Q}, either immediately -- for reads, or only once they are
successfully replicated -- for updates. 
The scheduler checks each operation against all currently
executing operations in the set {\tt E}. If it commutes with all operations in {\tt E}, the operation is added
to {\tt E} for execution. Operations at the head of {\tt Q} that have failed replication (perhaps due to a RAFT
leadership change after the operation was received) are removed from consideration by the scheduler.

\newlength\mylen
\newcommand\myinput[1]{%
	  \settowidth\mylen{\KwIn{}}%
		  \setlength\hangindent{\mylen}%
			  \hspace*{\mylen}#1\\}

\begin{algorithm}
  \caption{\sys operations scheduler algorithm.}
  \label{schedulealgo}
\SetKwInOut{Input}{input}  
\Input{\footnotesize {1. queue Q of updates and reads. \\
 2. Set E of operations that are ready-to-execute/executing \\}} 

\Repeat{server-shutdown}{
	\If{Q.head().state == FAILED\_REPLICATION} {
	         Q.dequeue()
	}
	\ElseIf{Q.head().state != REPLICATING and Q.head() commutes with all ops in E} {
	        op = Q.dequeue()\\
	        E.insert(op)
	}
}
\end{algorithm}

\subsection{Execution}
The execution of operations in \sys is done by dedicated executor
threads. Each thread repeatedly selects 
an operation from the set {\tt E},
executes it, and after the execution is completed and any of its effects
persisted to memory, it is 
removed from {\tt E}.  The execution follows the steps detailed in
Algorithm~\ref{execalgo}. 

\begin{algorithm}
  \caption{\sys Execution algorithm.}
  \label{execalgo}
\SetKwInOut{Input}{input}  
\Input{\footnotesize {o = operation to execute, E = set of executing/ready operations\\}} 
\If{o.state != COMPLETED} {
	 BEGIN FAILURE\_ATOMIC\_TX\\
	 remove o from E\\
	 lockset = []\\
	 HandleRPC(o.request, lockset, o.response)\\
	 o.state = COMPLETED\\
	 END FAILURE\_ATOMIC\_TX\\
	\ForEach{l in lockset} {
		 l.ReleaseLock()
	}
}
remove o from E\\
Send o.response to client
\end{algorithm}

We note that although the scheduler ensures only commutative operations  execute simultaneously, that does not
mean those operations will not conflict. 
For example, increments to a counter are commutative but
one still needs to synchronize on access to the counter to avoid two
updates reading the same initial value of the counter.
In addition, total order on simultaneously executing concurrent
updates is needed to make recovery possible.
Therefore we need to ensure commutative operations serialize with each other as a whole when accessing the same
memory location. We enforce this via delayed release of locks -- a fairly standard technique borrowed from databases.
The execution algorithm ensures that {\em any locks acquired during execution (by user code) are released only after execution
and persistence are complete}. 

We draw particular attention to recovery. The persistent circular log and state flags in the RAFT log entry form the
foundation for recovery. We process all
undo logs and then start the scheduler. The delayed lock release ensures 
that any operation that had completed does not see any change to its input data. If an operation begins execution but fails before executing, its
persistent state flag remains set at {\tt REPLICATED} when the system restarts. Any persistent changes
made by the previous execution are automatically undone by PMDK. It then proceeds as usual through the current
attempt till completion. On the other hand, if an operation finishes execution and manages to move its persistent
state flag to {\tt COMPLETED}, we do not execute it again by checking for this condition, thereby ensuring that
operations are executed exactly once with respect to changes to
persistent memory. 

\subsection{Commutativity}
\label{sec:commutativity}
We now consider how commutativity impacts the distributed consistency model. The precise definition
of commutativity that we provide to \sys programmers is: \emph{Two
  operations commute if the result 
returned by each of them, when executing one after the other, does not depend on the order of execution}. 
This relation is transitive. When no two operations are defined as commutative, execution occurs in RAFT log order and 
therefore we provide serializability with read-after-write consistency with respect to the data structure operations.
If two operations are declared to commute, \sys might execute them in different orders on different replicas
but, by definition, this cannot change their results and therefore does not cause a non-serial schedule to become visible.

To illustrate this, 
consider the case of two locations, A and B both initialized to 0 and both set to 1 by concurrent writes. 
The writes are commutative and can be reordered. If two different
clients try to read A and B, one can see the state
A==1 and B==0,
while the other sees the state
A ==0 and B==1. This is consistent with the following serialized schedule:

\noindent  Read1(A)=0; A:=1; Read2(A)=1; Read2(B)=0; B:=1; Read1(B)=1.


\noindent We reemphasize here that we provide serializability and not
linearizability.  In addition, commutativity constraints prevent reads from bypassing writes to the same location. This means on
a leader failover, committed writes in the log must be executed at the new leader before reads to the same location are allowed
to execute at it.

Commutativity can be a powerful tool to extract parallelism from the sequential order specified in the log.
As an example of allowing some commutativity, consider a single container in persistent memory with a dictionary
(implemented as a persistent hash table) interface. Most such APIs (e.g., in C++ STL containers) disallow operations
to multiple keys. Therefore a natural setting for commutativity is to allow operations (reads or writes) to different keys to commute,
since reads by clients cannot reveal out of order application of
operations to different keys at different replicas. In such a situation, the
programmer can set {\tt Commutes} to return true if and only if the operations are made to different keys. The result is
serializability with read your own writes consistency when the data structure API is \emph{restricted} to a single key.

As an example of a more complex commutativity specification consider an example of a graph stored in persistent
memory, as an adjacency list: a map of vertices to a list of neighboring vertices.  Adding or deleting edges can be tricky 
due to the need to update both source and destination vertices. We need to ensure that reads see a consistent state of the graph:
reading attributes of an edge specified as $(u, v)$ should succeed and return the same result regardless of whether we
lookup vertex $u$ or $v$ to retrieve edge information. An intuitive setup here is to allow edge changes to commute
if they do not touch the same vertex: $Commutes((u, v), (x, y))$ should return true if and only if
$\{u, v\} \cap \{x, y\}  = \emptyset$. 

We show in the evaluation that judicious settings
of commutativity allows more concurrency to be extracted from the single serial order in the RAFT log and therefore
better performance. We note that the assertion that APIs that allow more calls to commute lead to more
concurrency is in fact a \emph{general} notion~\cite{commuter,crdt}. 
Prior work~\cite{rinard1997commutativity} presents mechanisms to identify commutative 
operations from a programs exposing operations API (e.g., \sys data structures) using compiler support. 
They use data dependence analysis and symbolic execution on operations/input-objects to uncover API calls that commute with each other. 
Tooling support to automatically identify commutative rules yields more parallelism/better performance in \sys
usecases and also eliminates potential manual specification errors; we
leave the development of such infrastructure to future work.  


%% file: impl.tex
\section{Implementation Detail}

The implementation of the replication layer in \sys uses an open
source C/C++ library for the RAFT protocol~\cite{raft-impl}. With 800
lines of C/C++ code we extend
the RAFT library to directly use Intel's DPDK network transport -- a userspace networking stack that runs
on Ethernet network fabrics. 
The design of \sys, however, is applicable to other user-level network
stacks with direct memory access. 
In order to carry out the evaluation of \sys, we also implemented
TCP/IP transport extensions for the same library
(see~\S\ref{eval:replication}). 

The Blizzard optimizations related to the replication transport, such as zero-copying and batching, require greater
control over network packet processing. 
Using 
DPDK provides this control, however, to use it for building a {\em
  persistent} replication transport that works directly with PMEM,
requires solving two additional problems. First, DPDK as-is does
not interoperate with persistent memory allocators.
We solve this by modifying the DPDK allocator to obtain
2~MB chunks of memory from a user-defined DAX file path, using the standard
named \texttt{mmap} calls, supported by PMDK. 
This allows DPDK to allocate network packet buffers directly from the persistent memory.
Second, the allocated buffers are exported to applications as virtual
memory addresses that are not unique across restarts, and thus
cannot be used as a unique handle for persistent memory programming.
We solve this by durably recording
the mapping between the \texttt{mmapped} file name and the virtual
addresses at startup time, and use  the same mapping to
re-construct/re-map the old virtual addresses to new ones during subsequent restarts.  

These two solutions make it possible to implement a crash-consistent transport
using DPDK and persistent memory, and form the basis for the
implementation of \sys. 

%% file: evaluation.tex
\section{Evaluation}
\label{sec:eval}
\noindent{\bf Setup:} We evaluate \sys on a cluster of three identical servers  (\autoref{table:testbed}) and 
a set of client machines to issue RPC calls. The servers are connected to each other via a 10 GigE switch
with jumbo frames enabled. 
Clients talk to servers via the same Intel DPDK
network transport.
Each server node has DRAM and PMEM memory modules that are accessible via load/store instructions.
An \texttt{ext4} DAX file-system configured to use 2 MB hugepages~\cite{pmem-hugepages} manages the PMEM
address space.  Unless otherwise mentioned, we allocate message buffers (DPDK mbufs), 
RAFT log entries, persistent data-structures and their undo logs entries from PMEM. 

\begin{table}[t]
	\centering
	\footnotesize
	\begin{tabular}[htbp]{| c | c |}
		\hline
		Compute & \makecell[l]{Intel Xeon Cascadelake, 56 cores @ 2.2 GHz over \\ 2 NUMA sockets. Running Fedora Linux.}\\ \hline
		Memory & \makecell[l]{375 GB of DRAM memory and 756 GB of direct \\  accessible, PMEM memory side by side to each other} \\ \hline 
		PMEM & \makecell[l]{756 GB of Intel Optane Persistent Memory (200 series) \\ in side-by-side to DRAM (App Direct mode)} \\ \hline 
		Flash & \makecell[l]{420 GB Intel 520 series} \\ \hline 
		Network & \makecell[l]{Commodity Intel 10-Gigabit Ethernet network cards with \\ Intel DPDK userspace network stack.} \\ \hline
	\end{tabular}
	\caption{Node configuration in the \sys testbed}
	\label{table:testbed}
\end{table}

\subsection{Replication}
\label{eval:replication}

\begin{figure}[t]   
	\centering
	\includegraphics[width=\linewidth]{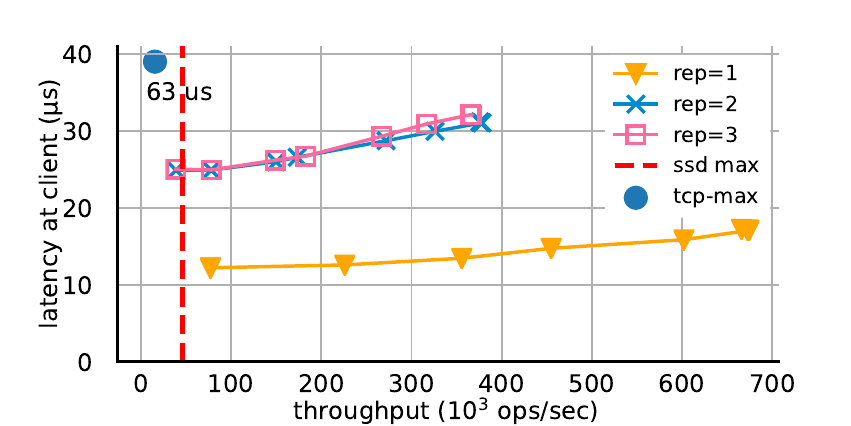}
        \vspace{-5.0ex}
	\caption{\small Replication performance characteristics in \sys for varying replicas. N in rep=N, represents 
	N-way replication. \texttt{tcp-max} and
	\texttt{ssd-max} represents the theoretical peak performance of a TCP/IP and SSD based replication system.} 
      \label{plot:echo}
      \vspace{-3.0ex}
\end{figure}

We evaluate the basic \sys replication performance using a no-op
microbenchmark where 
the leader returns a response to the client as soon as the log entry has been replicated to a majority of
the quorum of server nodes. \sys is configured to batch 32 log entries into a single RAFT log entry for efficiency. 
To put the results in perspective, we
run the same benchmark with a regular network transport by replacing
our userspace networking stack with TCP/IP, and also 
plot the theoretical peak performance
if we were to batch-write 32 log entries at a time to Flash storage. 

\autoref{plot:echo} demonstrates that the presence of PMEM removes 
storage bottlenecks and allows  replication to be unhindered by the cost of persistence. 
This necessitates an optimized implementation that reduces network overheads, 
so as to prevent the network stack from 
becoming a bottleneck. 
\sys is able to achieve this and provides a raw replication rate of $\sim$$365K$ 
log entries a second (3 ways with full persistence).

\begin{figure}[tb]   
	\centering
	\includegraphics[width=\linewidth]{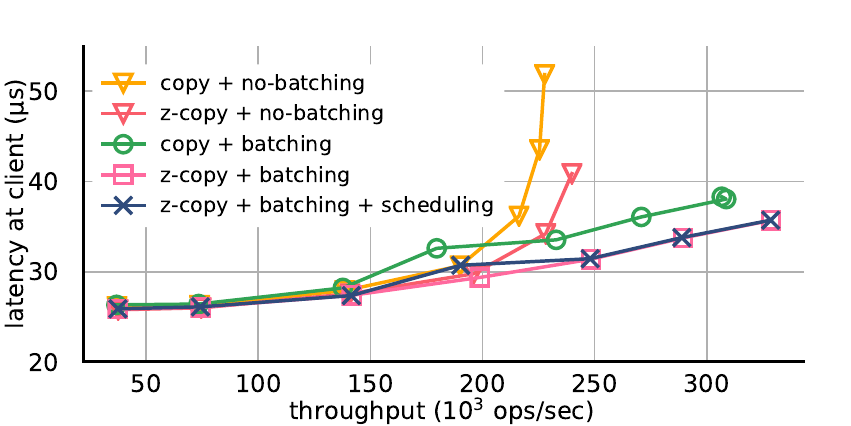}
        \vspace{-5.0ex}
	\caption{\small \sys's zero-copy (z-copy) and batching improve system latency and throughput. 
	The commutative scheduler only introduces negligible performance overhead.}
      \label{plot:optimizations}
      \vspace{-3.0ex}
\end{figure}

In~\autoref{plot:optimizations},  we show the gains of the
mechanisms integrated in \sys, using the same replication
microbenchmark in a system 1) without batching optimization (no-batching), where
we process one operation at a time during log-appends and replication and  2) without zero-copy 
optimization (copy), where we make copies of incoming RPC payload, both during RAFT log-appends
and data preparation during network multicast, as our baselines. 
Then we introduce zero-copy (z-copy) and batching optimizations into the system separately as well
as both together.

Zero-copy reduces the \sys latency at the clients by $\sim$$36\%$ at peak system throughput of 216K ops/sec.
Zero-copying eliminates extra memory management -- allocation, data-copy and free, during \sys's log persist and replication steps.
The optimization contributes to modest performance gains under current
DPDK based network stack, but is likely
to have even more significant impact on faster network stacks such as
RDMA, with an order of magnidue lower network latencies
($\sim$$1$us
latency 
for EDR InfiniBand RDMA vs.~$10+$us on our 10Gbps Ethernet
testbed).
The effects of zero-copy become more
significant at higher loads, because they eliminate contention for
PMEM accesses: PMEM throughput, particularly write throughput, is
known to degrade drastically with concurrent
operations~\cite{optane:perf}. 

Use of batching in \sys leads to 40\% 
higher throughput, compared to the 
baseline. 
We attribute the performance gains in batching to 
the increased memory parallelism during RAFT log appends, since a batched RAFT log entry append (for 32 ops) 
requires only a single store fence instruction after subsequent CPU cache line flushes, 
whereas the no-batching version requires 32 of them, and to 
the reduced network multicast cost, as the RAFT metadata appends and control path operations are amortized over
the number of batched operations.
Zero-copy and batching combined together enable \sys to handle peak
throughput of 328Kops and per replicated 
operation latency \textless\SI{36}{\micro\second}.

Zero-copy and batching improve replication throughput, but also
shift the system bottleneck from the replication layer, back into the persistent data-structures. 
The commute scheduler in \sys enables concurrent operation execution on data-structures
without compromising correctness. We use the microbenchmark to measure
the overheads of the scheduler invocation (+scheduler), without
actually changing the behavior of the executed operations. 
The result is close to one with no scheduler, demonstrating that the dominant cost in 
checking commutativity is on the side of the callback provided by the user, rather than the scheduler implementation.

\begin{figure}[tb]   
	\centering
	\includegraphics[width=\linewidth]{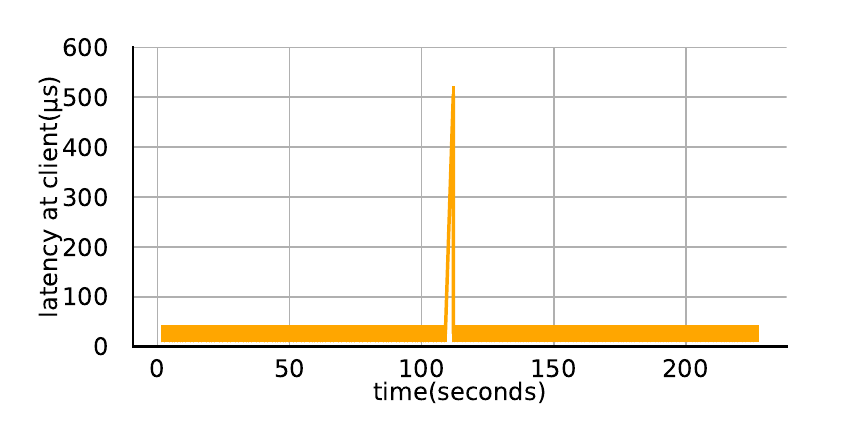} 
	\caption{\small \sys failover performance in a 3 node replica cluster. } 
	\label{plot:failover} 
\end{figure}

Finally, we show how \sys handles replica failures to provide applications with crucial availability
guarantees beyond raw performance improvements. In~\autoref{plot:failover} we show the failover timeline of a \sys
cluster with 3 nodes under echo/no-op workload, where we kill the
leader replica 
midway. After a timeout-based failed leader detection, the \sys client
probes the other replicas for a new leader.
For a failure detection timeout of 12ms, \sys fails over to a new
leader within 24 ms
in the worst case, 
for a 3 node replica cluster.

\subsection{Key Value store}
Persistent key-value (KV) stores such as RocksDB~\cite{rocksdb}, as an example of
general-purpose application often used as a persistence layer,
use complex data structures such as LSM trees~\cite{lsm}   
to compensate for the poor random access over block storage. 
The byte-addressability of PMEM makes it possible to build a simple
and performant
key value store
based on a hashmap (e.g., as done with
Persimmon~\cite{persimmon:osdi20}), which is the 
in-memory data-structure realization for the KV 
abstraction. 
We ask the following question: 
Is it feasible implement a truly persistent, replicated hashmap using the 
\sys programming model? 

We implement a hashmap-based key-value 
store that supports point queries. With $96$ lines of C++ code, we port
the concurrent  hashmap
implementation from PMDK 
as a \sys data-structure by extending it with \sys's crash-consistent
update protocol and a commute handler that honors a serialized read-your-own writes
consistency model, as outlined in \S\ref{sec:commutativity}. The implementation supports arbitrary
strings of characters as both keys and values.

We select as a baseline NoveLSM~\cite{novelsm} -- a KV store optimized
for PMEM. 
NoveLSM introduces to traditional LSM-tree based KV store designs
PMEM-optimized mutable memtables,
in-place updates, memory optimized storage transactions, and parallel reads. 
We place the mutable memtables and lower-level SSTs of NoveLSM on PMEM and run 
it on \sys with replication and commutative
scheduling turned off, thus \sys only serves as an RPC transport for NoveLSM. 
We use 8 byte key/value strings for implementations, 
and a Facebook-like~\cite{memcacheFB,fbworkload}
workload with 50\% writes and a uniform distribution of keys.
We use a write-intensive workload in the evaluation since this is when
it is critical to provide for 
scalable replication and persistence. 

\begin{figure}[tb]   
	\centering
	\includegraphics[width=\linewidth]{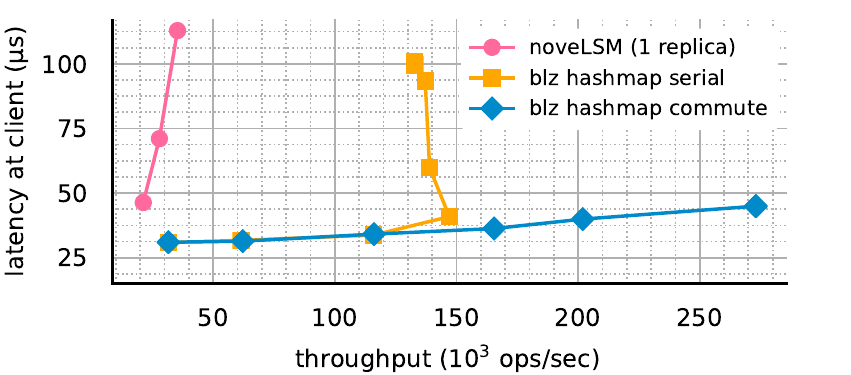} 
	\caption{\small Blizzard's key-value map performance comparison (replicated and persistent) against similar software stacks} 
	\label{plot:kv} 
\end{figure}

The results in \autoref{plot:kv} show that the 
replicated and crash-consistent concurrent hashmap-based key-value
store with \sys (blz hashmap serial) outperforms 
NoveLSM's (single replica) peak throughput by as much as $7\times$,
merely due to not being constrained to use block
interfaces and LSM trees.
Importantly, the throughput of \sys further improves by
a factor of two, up to a throughput of 270Kops,  
once we mark operations
to different keys as commutable (blz hashmap commute) thereby removing the constraint of serial execution imposed by the RAFT log.  
We did not see a significant increase in operation latency of our
hashmap, where it remained \textless\SI{44}~us across the throughput
range.
This result
shows that \sys allows
programmers to exploit this without being burdened by implementing
their own crash consistency or replication. It also 
underlines
the importance to leverage the byte-addressability of PMEM and the
network stack in the design of the replication layer, and of the
support to extract increased parallelism during replication via the
use of the commute API. 

\subsection{Graphs}
Next, we evaluate \sys in the context of  graph databases -- an important class of 
applications that presents unique durable data management challenges.
The most natural representation of graphs
for search and traversal uses pointers -- a source of great difficulty
with block interfaces, 
that has spawned a whole genre of research into
batch processing of graphs from secondary storage \cite{graphchi,
  x-stream, chaos}. 
%
PMEM, natively allows pointers and persistence to co-exist and
presents a unique opportunity. 
The key challenge is to do this without letting persistence or
replication add programming complexity or  unduly affecting 
performance. We examine whether
\sys is up to this task.

An adjacency-list is the natural data-structure for representing graphs on memory. 
We first implement a persistent adjacency list data-structure by putting together
already available PMDK building blocks. Persistent list structures contain neighbor lists
and a hashmap structure maps a vertex's 
\texttt{\small node\_id} to corresponding \texttt{\small list\_entry} of the row. We 
extend the implemented graph-structure with \sys's crash-consistency
semantics. Finally,
we implement the handler for parallelizing commutative operations. 
The implementation only took 110 lines of C++ code, as the bulk of the building blocks were
already available as open source libraries.

We compare our implementation with  the GraphOne~\cite{graphone}, an in-memory graph processing system.
GraphOne~\cite{graphone} models a graph using combination of an in-memory edge-log (memory-buffer), an in-memory versioned adjacency-list and a 
persistent edge-log.
The incoming graph updates are first buffered in the memory-buffer before simultaneously moved/archived into the versioned, in-memory 
adjacency-list and the file-backed persistent edge-list. Archive
happens after configurable number of edge updates and read requests on
a 
graph are served from the in-memory adjacency-list that is up-to-date till the last archival epoch. The configuration is known as
\texttt{\small static-view} with \texttt{\small stale-reads} in GraphOne terminology
and trades off the read freshness for speed.
We run GraphOne as a network service, using \sys's RPC transport. We
place the GraphOne durable edge-log on a PMEM backed file-system 
and use \texttt{\small static-view} API with \texttt{\small stale-reads}. The
archival epoch is set to be once per every $2^{12}$ edge inserts.

We use the Twitter data-set~\cite{twitter-paper} as our streaming graph benchmark. The data-set includes 
a subset (up to 15M nodes, 46M edges) of Twitter social network in the form of who-follows-who. 
We use half of the data-set to pre-load our
graph database and use the rest of the updates to form a read/write (50/50) streaming workload similar to~\cite{graphone}.
We read the out-degree for a given node, as our read operation in this workload. 

\begin{figure}[tb]   
	\centering
	\includegraphics[width=\linewidth]{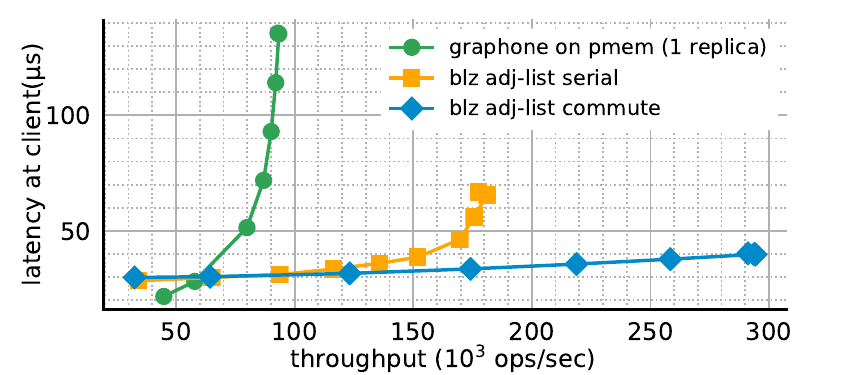}
        \vspace{-3.0ex}
	\caption{\small Twitter benchmark against GraphOne and \sys's adjacency-list based graph engines} 
	\label{plot:twitter}
         \vspace{-3.0ex}
\end{figure}

\autoref{plot:twitter} shows that 
the \sys graph representation based on a persistent adjacency list (blz adj-list serial) can handle 
up to 150Kops, at \textless\SI{38}{\micro\second}, 
all while providing strong (read your own writes) reads and full fault
tolerance semantics. GraphOne
only manages a peak throughput of 80Kops and takes as much as \SI{50} {\micro\second} for the same 
workload. 
GraphOne integrates extra versioning techniques into its in-memory adjacency list to decouple in-memory operations
from persistent write overheads. However, the software overheads introduced by these extra steps dominate when presented with 
PMEM based fast persistence. 
We disabled the persistent semantics of the GraphOne engine and ran the same experiment and the results (not shown for clarity) confirmed
our reasoning.


Parallelizing the adjacency-list representation by specifying commutativity
rules further improves the \sys 
graph operation throughput to 291Kops, a $\sim$$2\times$ improvement over a strictly serial execution schedule without
significant increase in operation latency. Overall, the \sys replicated and fault-tolerant adjacency-list based native
graph representation outperforms the GraphOne graph engine by $\sim$$3.6\times$, even without exotic data-structure level
optimizations.

\subsection{Lobsters}

Finally, we use \sys to implement and evaluate a persistent data storage backend for a popular 
web-application, Lobsters~\cite{lobsters}. Lobsters is a community based
news aggregation site where users vote for submitted web-links. They display the 
top-K voted web-links on their home page. The original site manages 
web-links and their vote counts using a relational DB backend~\cite{noria}. They model durable state
using \texttt{\small article(article\_id, web-link,...)} and 
\texttt{\small vote(article\_id, vote\_count)} relations. An article submit inserts a new entry 
into \texttt{\small article} relation and an upvote/downvote updates
the \texttt{\small vote} relation.
The top-K voted article list is maintained as application logic.

\begin{figure}[tb]   
	\centering
	\includegraphics[width=\linewidth]{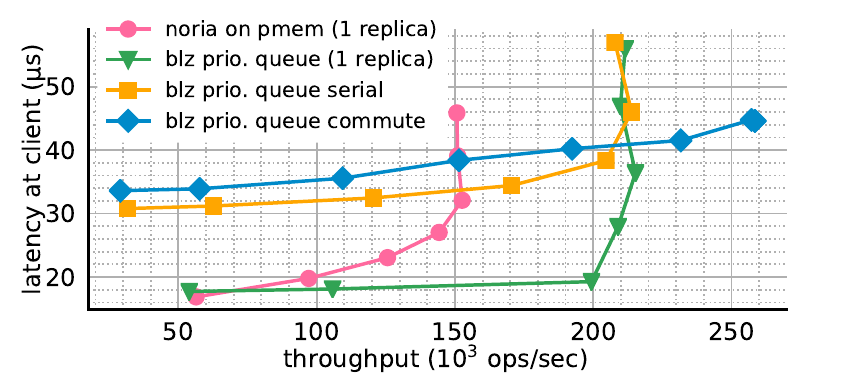}
         \vspace{-3.0ex}
	\caption{\small Lobster vote benchmark performance numbers for Noria and \sys persistent priority queue based storage backends.} 
	\label{plot:vote}
         \vspace{-3.0ex}
\end{figure}

We use a persistent priority queue as the memory native data-structure 
to serve top-K requests. Such a data-structure is both intuitive and removes auxiliary
book keeping in the form of application's runtime state as the data-structure itself maintains
ordering with updates in logarithmic time. We use a max-heap to maintain the top-K voted
stories and a min heap for the remaining stories. An upvote potentially causes the most voted
story in the min-heap to move to the max-heap displacing the story with the minimum votes in
the current top-k. A downvote can cause the opposite to happen. The hashmap 
maps \texttt{\small article\_id}s to min/max-heap entries as the incoming vote requests 
are indexed using \texttt{\small article\_id}. For parallelism, we use a sharding scheme 
with multiple min-max heap pairs and shard the story keyspace across them. Determining
the top-k becomes slightly more expensive due to the need to combine the results of the union of the
min-heaps. We combine k entries from each sharded priority queue to form the final
result. 

We re-used a persistent hashmap implementation from PMDK~\cite{pmdk} and implemented a 
persistent top-k priority queue using newly written min/max-heap code.  Implementing the
core persistent data-structure and operations took $\sim$$600$ lines of C++ code. We only allow
update operations to different shards to commute with each other.

We compare \sys's fault tolerant priority-queue data-structure performance
against a Noria~\cite{noria} based Lobsters backend that  represents
that bestcase Noria performance. 
The authors of Noria obtain their bestcase performance using RocksDB. 
We upgrade the choice of KV store to NoveLSM, as it is optimized for
PMEM~\cite{novelsm}. 
Both \texttt{\small article\_id} to \texttt{\small title} and
\texttt{\small article\_id} to \texttt{\small vote\_count}
mappings are encoded as key-value strings.  
For a key-value store backend a top-K operation would be very costly as a join routine needs to be performed external to the 
data store (e.g., in application logic). Therefore, following the original Noria benchmarking setup~\cite{noria} for KV-stores, we convert the top-K 
requests to simple reads, thus pushing the Noria-like backend to its best case performance.

We run the Vote benchmark from \cite{noria}: a zipfian load generator that is modeled after actual
Lobster website traffic. We preload the data store with 1M articles and run the experiment with 19/1, read/update
traffic.
The update operations consist of up-votes/down-votes of
articles. During a read, the Noria based Lobster backend simply
returns the article information for a given \texttt{\small article\_id},
whereas the priority-queue based \sys backend returns the top-K
articles at a given instance. Therefore a top-K read request with the
\sys implementation on average moves K$\times$ more data than the Noria
counterpart. We use K=8 over a 4-way sharded priority queue and issue
one write operation for every 20 request. 

\autoref{plot:vote} shows that 
the Noria-based Lobsters backend serves up to 150Kops at 
\textless\SI{30}{\micro\second}. The \sys priority-queue based backend (blz prio. queue serial) handles
peak throughput of $\sim$$200$Kops but incurs \textless\SI{55}{\micro\second} per operation due to 3-way replication.
It is important to note that the Noria best case performance numbers benefit
from the relatively
simple read workloads (no top-K) and lack of fault tolerance
(no-replication) over \sys.
The sharded and parallelized version of the \sys persistent priority
queue 
(bliz prio. queue commute) 
manages a maximum throughput of $\sim$$257$Kops while keeping operation latency under \textless\SI{45}{\micro\second}. 
Data structure sharding along with proper parallelization of incoming operations using commutativity helps \sys data structure to 
handle 70\% more traffic compared to Noria, while maintaining the same intuitive memory APIs.

%% file: relatedwork.tex
\section{Related Work}

Over the years the systems community has worked on several key
software components
for PMEM. 
PM-aware file systems~\cite{pmfs,bpfs} provide direct-access (DAX) and enable memory 
native PM programming for userspace applications. Failure-atomic PM programming primitives
\cite{wblogging, justdologging, dudetm}, 
and PM-friendly synchronization primitives~\cite{atlas,nvthreads,mnemosyne,dhtm,nvheaps}
enable both easy and correct PM programming.  
These lines of studies have  collectively
contributed towards persistent allocators~\cite{pmdk,makalu}, storage engines~\cite{aerie,novelsm}
and data-structures~\cite{bwtree,persimmon:osdi20,pronto}. However
these solutions have limited reliability and  
availability as the the persistent media is vulnerable to node
failures.

Among these prior efforts, Persimmon bears closest similarity with
Blizzard's offering of persistent data structures as a
service~\cite{persimmon:osdi20}. Unlike \sys, Persimmon guarantees
linearizability for the persisted state, but at the cost of supporting
only single threaded applications and enforcing sequential log
updates. 
In addition, log replication is used only
to capture changes of in-DRAM data structures and persist them to
PMEM, thus it does not offer true persistence.  
Naively adding replication to this solution will incur
copying and serialization overheads, and fall
short on the performance efficiencies that \sys extracts through its
coupled log and commutative API.

Maintaining distributed consistency among replicated data while maintaining good system performance
is one of the key challenges in the design of reliable 
system. Some system designs rely on highly available
external coordination services~\cite{zookeeper,chubby} to maintain consistent replica state.
Other system designs~\cite{bolosky2011paxos,malkhi2012paxos} 
opt for a self-sufficient deployment model where the nodes
handle both replication and coordination themselves without external
co-ordinator.
CRDT like system abstractions~\cite{crdt}  guarantee replica state
convergence using restrictive APIs. 
The commutativity between operations guarantees eventual convergence without depending on the 
operation ordering.
As these systems were using block storage for durability, the trade-offs that drove
their designs are outdated in the era of PM programming.
\sys also bears similarity to 
other work which exploits commutativity to improve
replication~\cite{curp:nsdi19}.  CURP~\cite{curp:nsdi19} does not
  consider PMEM, instead clients assume responsibility to replicate data to
  all replicas where data is immediately persisted, but 
  application commit is immediate only for commutative
  operations, otherwise a consistent ordering must first be 
  established using RAFT. \sys assumes only RPC as a client
  API, and aims to maximize the overlap of the presistence- and
  replication-related operations.



Recent work on PM has explored the feasibility of supporting reliable persistent memory programming.
Mojim~\cite{mojim} provides reliable, PM aware data storage to application by overloading 
\texttt{\small mmap/msync} system calls with synchronous replication
support. Mojim uses an external 
co-ordinator and only supports strong replication 
to one mirror node 
in its most common deployment mode.
AsymNVM~\cite{asymvm} has a client-server model for PM programming similar to \sys. AsymNVM uses
both value and operation logging for efficient log replication and
features an LRU cache in the client
for fast reads.
However, its replication mechanism is outside of the critical path of
persistent writes to (remote) PMEMs, and relies on an external
coordinator (Zookeeper) 
for consistent state management.
This is appropriate for this system, since replication here assumes
mirror nodes may be arbitrary persistent device, including much slower SSD devices. 
In contrast 
\sys uses a unified replication log for both consensus and operations
tracking that allows it to provide true persistence while maintaining
performance. 


\section{Conclusion}
In this paper we
consider support for reliable and persistent programming 
in the context of modern persistent main memory and fast networking
hardware. We build \sys, a reliable, persistent system software stack modeled
after main memory programming. \sys allows developers to program their
reliable, durable application state using familiar main memory
data-structure abstractions.
Designing and implementing \sys involved solving two main
challenges.
To keep up with PMEM speeds required high-performance replication with
fast user-level replication. For maximum performance, and to fully
leverage the byte-addressability of PMEM, required tight integration
of replication and PMEM programming abstractions. By exposing to
clients an API centered around data structure operations, allowed \sys
to leverage log replication and integrate correctness protocols for
failure atomicity and commutativity (i.e., ordering).
Using three real world application workloads with \sys-powered
backends, we show
\sys can realize persistence and fault tolerance
with only modest software changes and with significant performance
gains.

